\documentclass[runningheads]{llncs}
\usepackage[T1]{fontenc}
\usepackage{graphicx}
\begin{document}
\title{CertiA360: Enhance Compliance Agility in Aerospace Software Development}
\author{J. Antonio Dantas Macedo\inst{1}\orcidID{0009-0001-7168-2569} \and
Hugo Fernandes\inst{}\orcidID{0000-0002-2135-4110} \and
J. Eduardo Ferreira Ribeiro\inst{1}\orcidID{0000-0002-1894-3993}}
\authorrunning{José Dantas Macedo et al.}
\institute{Department of Informatics Engineering, Faculty of Engineering, University of Porto, Porto, 4200-465, Portugal \\
\email{up201705226@up.pt}}

%\institute{Department of Informatics Engineering, Faculty of Engineering, University of Porto, Porto, 4200-465, Portugal}

%\author{José Dantas Macedo\inst{1}\orcidID{0009-0001-7168-2569} \and
%J. Eduardo Ferreira Ribeiro\inst{1}\orcidID{0000-0002-1894-3993} \and
%Ademar Aguiar\inst{1,2}\orcidID{0000-0002-4046-4729}}
%
%\authorrunning{José Dantas Macedo et al.}
%\institute{Department of Informatics Engineering, Faculty of Engineering, University of Porto, Porto, 4200-465, Portugal \\
%\email{up201705226@up.pt} \and
%INESC TEC, Porto, 4200-465, Portugal}

%
\maketitle              % typeset the header of the contribution

\begin{abstract}
Agile methods are characterised by iterative and incremental processes with a strong focus on flexibility and accommodating changing requirements based on either technical, regulatory, or stakeholder feedback.
However, integrating Agile methods into safety-critical system development in the aerospace industry presents substantial challenges due to its strict compliance requirements, such as those outlined in the DO-178C standard.
To achieve this vision, the flexibility of Agile must align with the rigorous certification guidelines, which emphasize documentation, traceability of requirements across different levels and disciplines, and comprehensive verification and validation (V\&V) activities.
The research work described in this paper proposes a way of using the strengths of the flexible nature of Agile methods to automate and manage change requests throughout the whole software development lifecycle, ensuring robust traceability, regulatory compliance and ultimately facilitating successful certification.
This study proposes CertiA360, a tool designed to help teams improve requirement maturity, automate the changes in traceability, and align with the regulatory objectives. The tool was designed and validated in close collaboration with aerospace industry experts, using their feedback to ensure practical application and real-life effectiveness.
The feedback collected demonstrated that the automation given by CertiA360 may reduce manual effort and allow response to changing requirements while ensuring compliance with DO-178C. 
While the tool is not yet qualified under DO-330 (Tool Qualification), findings suggest that when tailored appropriately, Agile methods can not only coexist with the requirements of safety-system development and certification in highly regulated domains like aerospace, but also add efficiency.

\end{abstract}

\section{Introduction}

Safety-critical software systems, particularly those in the aerospace domain, are governed by standards such as DO-178C (Software Considerations in Airborne Systems and Equipment Certification) to ensure reliability and safety \cite{DO-178C}. Ensuring compliance with these standards requires rigorous traceability and extensive documentation, making achieving successful certification challenging and resource-intensive, particularly concerning verification and validation (V\&V) efforts. While traditional plan-driven methods, such as the traditional Waterfall method, include mechanisms that ease compliance through structured and rigorous steps, they often lack the flexibility required to accommodate evolving requirements and dynamic project demands. In recent years, Agile methods have significantly transformed software development practices by emphasizing adaptability, collaboration, and iterative processes. These principles have demonstrated considerable efficacy in fostering innovation, enhancing team responsiveness, and facilitating incremental value deliveries. However, the inherent flexibility of Agile methods introduces substantial challenges when applied to safety-critical domains, where strict regulatory compliance and rigorous processes are paramount. This study addresses this fundamental dichotomy: how to reconcile Agile's dynamic practices with the requirements of stringent standards such as DO-178C. DO-178C, as a cornerstone standard in aerospace software development, provides comprehensive guidelines to ensure safety and reliability. Key challenges associated with achieving compliance include (i) DO-178C necessitates comprehensive documentation across all software lifecycle stages, potentially conflicting with Agile's emphasis on reducing process overhead; (ii) ensuring bidirectional traceability between requirements, design, and implementation, which is often labor-intensive and susceptible to human error without appropriate tools; and (iii) producing exhaustive evidence of compliance, which can introduce delays and increase costs, adding further pressure to resource-intensive projects \cite{Framework2023}. Although traditional plan-driven methods have demonstrated efficacy in ensuring compliance, they frequently encounter difficulties maintaining pace with evolving requirements and technological advancements \cite{Ribeiro2023DO,Framework2023}. Conversely, Agile methods, widely recognized for their iterative and flexible approach, encounter obstacles in fulfilling the exhaustive documentation and traceability requirements mandated by DO-178C \cite{SilvaCardoso2022,Framework2023}.
To address these challenges, this study proposes a solution, CertiA360, a tool designed to facilitate the integration of Agile methods with safety-critical compliance processes. We defined the following research question (RQ) to guide the development of our proposed tool, CertiA360: 

\textbf{How can a tool ensure automated traceability, manage change requests, and streamline documentation workflows in safety-critical software systems?} 

This paper is structured as follows. Section \ref{Background} provides an overview of Agile methods, safety-critical systems, aerospace software development, and DO-178C standard. Section \ref{RelatedWork} reviews related work and identifies the existing gaps in the field. Section \ref{ProposedSolution} introduces the CertiA360 tool and elucidates its architecture, features, and capabilities for addressing compliance and traceability challenges. Section \ref{ImplementationAndValidation} discusses the methodology used to develop CertiA360. Finally, Section \ref{Conclusion} summarizes the key findings and suggests directions for future research.

\section{Background}
\label{Background}

\begin{figure}[htbp]
\centering
\includegraphics[width=0.8\textwidth]{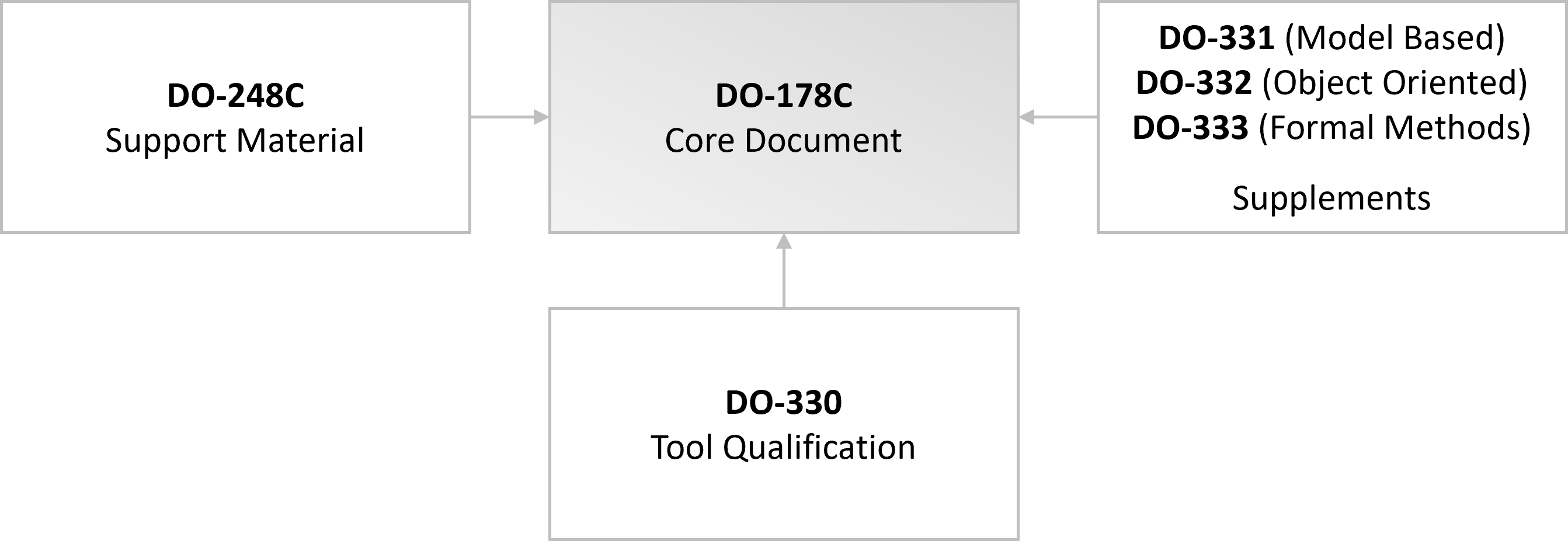}
\caption{DO-178C and related documents, adapted from \cite{Ribeiro2023DO,Ribeiro2023}.}
\label{fig2}
\end{figure}

Agile methods, safety-critical systems, and stringent compliance requirements of standards, such as DO-178C, intersect in complex ways, posing challenges to traditional development practices. Understanding these intersections is key to addressing gaps in the tools and processes for managing change, ensuring traceability, and achieving regulatory compliance. \textbf{Agile methods} emphasize iterative progress, collaboration, and responsiveness to change. As outlined in the Agile Manifesto \cite{AgileManifesto}, these principles prioritize incremental software delivery, interactions over processes, and adaptability to evolving requirements \cite{Ribeiro2023}. Scrum and Extreme Programming (XP) frameworks offer structured approaches to Agile implementation through sprints, user stories, and feedback loops \cite{ERTS2018}. Balancing the Agile's flexibility with compliance requirements such as those mandated by DO-178C is difficult \cite{Ribeiro2023DO,SilvaCardoso2022}. \textbf{Safety-critical systems}, such as those in aerospace, automotive, and medical devices, demand high reliability to prevent catastrophic consequences, such as loss of life or environmental damage \cite{Innovation2017}. These systems are governed by stringent regulatory standards that require thorough risk assessments, rigorous testing, and comprehensive documentation to meet the safety objectives\cite{Ribeiro2023}. \textbf{DO-178C} emphasizes lifecycle processes such as planning, development, verification, configuration management, and quality assurance \cite{DO-178C,Innovation2017} demanding extensive documentation and traceability. This often conflicts with Agile's preference for lightweight and flexible processes \cite{SilvaCardoso2022}. Bidirectional traceability is key, linking high-level requirements to implementation and testing artifacts. This ensures all functions are accounted for and verified, meeting safety-critical objectives. DO-178C categorizes software by Development Assurance Levels (DALs), with higher levels, such as DAL A, necessitating more rigorous documentation and validation \cite{Ribeiro2023DO,QScrum2020}. While essential for safety, these workflows can conflict with Agile principles \cite{SilvaCardoso2022}. The supplementary DO-331 standard specifies model-based development practices and streamlines compliance activities through automation \cite{A_Lean_and_HighlyAutomated}. Fig.~\ref{fig2} illustrates the relationships between DO-178C and its supplementary documents, including DO-248C, which provides additional guidance, and DO-331, DO-332, and DO-333, which address model-based, object-oriented, and formal methods, respectively \cite{Ribeiro2023DO,Ribeiro2023}.

In summary, tools such as DOORS and Enterprise Architect support requirement management in safety-critical software development, while systems such as Jira enable task tracking and sprint planning. Version control systems, including Git, effectively manage source code changes \cite{Innovation2017}. However, integrating these disparate tools into a cohesive framework remains challenging \cite{SilvaCardoso2022}. Current practices often rely on manual processes for traceability and compliance, which are time-consuming and error-prone. The growing complexity of modern systems and rapid technological advancements have challenged traditional methods, which require innovative approaches to balance compliance and adaptability \cite{SilvaCardoso2022}. Recent advancements in automated documentation and traceability management, such as those discussed in \cite{Framework2023} and \cite{SilvaCardoso2022}, offer promising approaches by embedding compliance mechanisms into Agile workflows.

\section{Related Work}
\label{RelatedWork}

Developing safety-critical software requires balancing stringent regulatory demands with the flexibility to adapt to technological advancements. Existing research explores different aspects of integrating Agile methods into safety-critical development and the role of document automation in addressing compliance challenges.

\textbf{Automation and Traceability}: traceability is a fundamental requirement in safety-critical software development, as mandated by standards such as DO-178C~\cite{DO-178C}. This ensures that every requirement is linked to design, implementation, and verification artifacts, validating safety objectives across the lifecycle. However, achieving bidirectional traceability remains labor-intensive and error-prone~\cite{Ribeiro2023DO,Framework2023}. Automation offers significant potential for improving traceability management. Tools that generate traceability matrices and validate coverage can reduce manual effort and enhance reliability~\cite{Ribeiro2023,SilvaCardoso2022}. Despite these advances, studies such as \textit{Escape the Waterfall} highlight the difficulty of integrating these tools into Agile workflows while maintaining compliance~\cite{Escape_the_waterfall_Ag}. 

\textbf{Managing Change Requests}: effectively managing change requests is another critical challenge in safety-critical domains. Agile methods provide opportunities to address changes during development~\cite{ERTS2018}. Practices such as continuous integration and automated impact analysis have shown promise in facilitating compliance within iterative environments by independently validating and documenting changes~\cite{Framework2023}. Nevertheless, Marsden et al.~\cite{ERTS2018} emphasized the ongoing challenge of aligning these approaches with stringent requirements of standards such as DO-178C. 

\textbf{Streamlining Documentation Workflows}: documentation plays a dual role in safety-critical software development, serving as a compliance artifact and a communication tool~\cite{Ribeiro2023DO,SilvaCardoso2022}. Traditional documentation workflows, often rooted in the Waterfall method or V-model, struggle to align with Agile's iterative nature~\cite{Innovation2017}. Emerging approaches like DocOps aim to integrate documentation into development pipelines, enabling real-time updates and reducing manual effort~\cite{SilvaCardoso2022,QScrum2020}. 

The literature review reveals several key insights such as: (i) automated traceability tools reduce manual effort and enhance compliance but are not yet fully integrated into Agile workflows~\cite{Ribeiro2023,SilvaCardoso2022}; (ii) effective change request management depends on automated validation and traceability mechanisms to uphold compliance with safety-critical standards~\cite{ERTS2018,Framework2023}; and (iii) documentation workflows need automation, yet current solutions often fail to meet safety-critical certification requirements~\cite{QScrum2020,Ribeiro2023DO}. In summary, addressing these gaps motivates the development of new approaches to address these challenges.

\section{Proposed Solution}
\label{ProposedSolution}

To address some of the challenges of integrating Agile methods into safety-critical software development, this study proposes \textit{CertiA360}, a centralized framework designed to harmonize Agile workflows with compliance requirements, ensuring traceability, adaptability, and effective collaboration across multidisciplinary teams. Its modular and scalable architecture allows it to seamlessly integrate existing tools while meeting project-specific regulatory needs.

CertiA360 acts as a third-party integrator connecting commonly used tools across safety-critical software projects. These include: (i) \textbf{Requirements management systems}: e.g., DOORs, Enterprise Architect; (ii) \textbf{Project management platforms}: e.g., Jira; (iii) \textbf{Version control systems}: e.g., Git. The tool parses and maps data from these systems into a centralized database, creating a unified platform where all stakeholders can visualize project information. This approach ensures consistency, transparency, and alignment with compliance standards such as DO-178C \cite{DO-178C}. Furthermore, CertiA360 is configurable to accommodate diverse regulatory requirements, making it adaptable to various safety-critical domains \cite{Framework2023}.

CertiA360's key features are aimed at addressing some of the complexities of safety-critical Agile projects:

\begin{enumerate}
    \item \textbf{Change Request Management} \\
    CertiA360 allows users to create change requests that automatically generate a comprehensive map of the required changes across all the affected artifacts. 
    % This feature ensures full traceability and provides a clear implementation roadmap, addressing the labor-intensive nature of manual compliance workflows \cite{Framework2023}.

    \item \textbf{Baseline Creation and Configuration Indexing} \\
    This tool enables users to create baselines for specific versions or releases, capturing the state of requirements, code, and backlog. This functionality consolidates all project artifacts into a configuration index, supporting accurate documentation and robust version control. 
    % Such approaches align with recommendations for ensuring auditability and compliance in iterative workflows \cite{SilvaCardoso2022}.

    % \item \textbf{Dashboards and Key Performance Indicators (KPIs)} \\
    % CertiA360 offers intuitive dashboards that provide insights into project data and track key performance indicators. This feature helps teams manage progress and identify bottlenecks effectively. Dashboards inspired by Agile methods, such as Scrum, provide real-time visibility into traceability and compliance status \cite{QScrum2020}.

    \item \textbf{Traceability Visualization} \\
    The tool includes graphical traceability visualizations, allowing users to filter and explore the relationships among requirements, code, tests, and other artifacts. This functionality enhances understanding and ensures compliance with traceability requirements outlined in DO-178C and DO-331 \cite{DO-178C,A_Lean_and_HighlyAutomated}.
\end{enumerate}
CertiA360 differentiates itself by bridging the gap between Agile practices and safety-critical compliance. It supports Agile workflows while maintaining alignment with regulatory standards, such as DO-178C, offering the following unique benefits: (i) \textbf{Regulatory Adaptability}--Configurable rules allow the tool to align with project-specific regulatory requirements, enabling compliance without compromising Agile principles; (ii) \textbf{Centralized Collaboration}--CertiA360 acts as a communication hub, providing updates and facilitating seamless collaboration among multidisciplinary teams; (iii) \textbf{Automation}--The tool automates labor-intensive tasks such as traceability mapping and change impact analysis, reducing manual effort and minimizing errors.

CertiA360 is designed to streamline communication and collaboration across large, multidisciplinary teams, addressing common challenges in Agile environments: (i) \textbf{Bug Resolution Workflow}--When a defect is identified, CertiA360 updates the affected requirements and artifacts, ensuring that all relevant teams are notified and can act efficiently. This workflow preserves traceability and compliance, as mandated by standards such as DO-178C \cite{DO-178C}; (ii) \textbf{Cross-Team Transparency}--The tool provides a centralized platform for accessing updated project information, reducing communication delays, and ensuring team alignment. 

By enabling efficient workflows and clear communication, CertiA360 ensures that Agile practices thrive even in highly regulated environments. 
% Its design and functionality balance adaptability and compliance, addressing the gaps identified in current solutions for safety-critical software development \cite{Framework2023,Ribeiro2023DO}.

\section{Development Process and Tool Architecture}
\label{ImplementationAndValidation}

\begin{figure}
    \centering
    \includegraphics[width=0.55\linewidth]{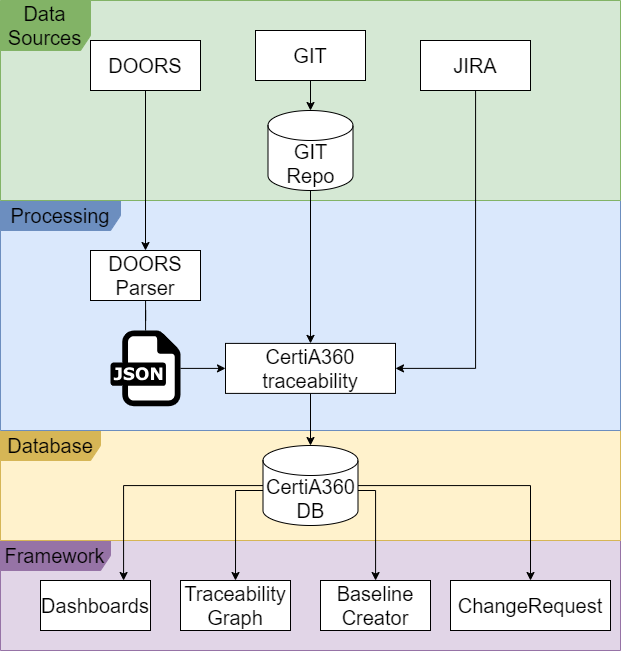}
    \caption{CertiA360 Architecture Layers}
    \label{fig:CertiA360-Architecture-Layers}
\end{figure}

The development process of CertiA360 was driven by an iterative and user-centered design approach, aiming to align with the requirements of safety-critical domains while adhering to Agile principles. Currently, CertiA360 exists as a proof-of-concept, and its architecture is designed to demonstrate its feasibility and foundational capabilities. Although not yet complete, its modular and scalable design provides a roadmap for future development and integration. The tool architecture comprises four primary layers, as shown in Fig.~\ref{fig:CertiA360-Architecture-Layers}.
\begin{enumerate}
    \item \textbf{Data Source Layer:} This layer facilitates seamless integration with external tools such as DOORS, JIRA, and Git. Leveraging robust API connections ensures that data from requirement management systems, project management platforms, and version control systems are accurately imported and synchronized.

    \item \textbf{Processing Layer:} The core functionality of CertiA360 resides in this layer. It includes modules for parsing, validating, and transforming data to ensure consistency and compliance with regulatory standards such as DO-178C. The validation module verifies the integrity of the imported data and identifies gaps or discrepancies in the traceability.

    \item \textbf{Centralized Storage Layer:} This layer provides a unified repository for all project artifacts.
    % The database schema is designed to maintain traceability links between requirements, code, tests, and other artifacts, ensuring data consistency across the development lifecycle.

    \item \textbf{Framework Layer:} This layer provides the key functionalities of the tool, including change request management, baseline creation, and traceability visualization. 
    % These functionalities are implemented through dedicated modules configurable to meet the specific needs of various safety-critical domains.
\end{enumerate}
% The modularity of this architecture enables CertiA360 to adapt to diverse project needs while maintaining scalability and performance. Furthermore, its API-driven design ensures compatibility, allowing teams to integrate the tool seamlessly into their existing workflows.

\section{Conclusion and Future Work}
\label{Conclusion}
CertiA360 demonstrates the potential to bridge Agile methods and safety-critical compliance. It offers a structured, automated approach to documentation, which enhances adaptability and efficiency. In addition to being part of a broader research initiative to enable continuous certification, this work lays the groundwork for further innovation in developing safety-critical software systems. Specifically, it contributes to significant advancements in the aerospace domain by aligning with and ensuring compliance with the DO-178C standard. This compliance enhances the reliability and safety of software systems and fosters the broader adoption of Agile practices with rigorous practices across the field.

Regarding the execution of real-world industry case studies to validate the tool results, ongoing research efforts are currently being applied to real aerospace projects. These projects serve as case studies for evaluating the proposed tool and validating its automation mechanisms. By focusing on the automation of output generation and reuse, this work not only enhances current processes but also lays a robust foundation for future improvements in efficiency and adaptability across various safety-critical industries. To advance this research, we have identified three primary challenges to be addressed in future work: (i) Conduct real-world industry case studies to validate the tool's results and effectiveness; (ii) Expand the validation scope to other safety-critical domains, such as railway, automotive, and medical devices, along with their respective regulatory standards; (iii) Continuously incorporate industry feedback to evolve the tool by integrating advanced features, such as predictive analytics, to enhance its utility and adaptability. In summary, it is imperative to maintain a strong relationship and collaboration between the scientific community and the industry. This partnership ensures access to industry data collection and analysis, which is typically unavailable in most research efforts because of confidentiality concerns. 

\begin{credits}
\subsubsection{\ackname} The authors thank Sérgio Santos and Marco Várzea for their continuous support and availability as aerospace and \emph{DO-178C} specialists.

\subsubsection{\discintname}
The authors have no competing interests to declare that are relevant to the content of this article.
\end{credits}


\begin{thebibliography}{8}

\bibitem{AgileManifesto}
Beck, K., Beedle, M., van Bennekum, A., Cockburn, A., Cunningham, W., Fowler, M., Grenning, J., Highsmith, J., Hunt, A., Jeffries, R., Kern, J., Marick, B., Martin, R. C., Mellor, S., Schwaber, K., Sutherland, J., Thomas, D.:
Manifesto for Agile Software Development. Agile Alliance. (2001).

\bibitem{DO-178C}
RTCA: DO-178C, Software Considerations in Airborne Systems and Equipment Certification. RTCA (2011).

\bibitem{Ribeiro2023DO}
Ribeiro, J., Silva, J., Aguiar, A.: Beyond Tradition: Evaluating Agile Feasibility in DO-178C for Aerospace Software Development. ArXiv Preprint arXiv:2311.04344 (2023).

\bibitem{Ribeiro2023}
Ribeiro, J., Silva, J., Aguiar, A.: Weaving Agility in Safety-Critical Software Development for Aerospace: From Concerns to Opportunities. IEEE Access, vol. 12, pp. 52778–52802 (2024).

\bibitem{SilvaCardoso2022}
Silva Cardoso Rodrigues, J.M., Ribeiro, J.E., Aguiar, A.: Improving Documentation Agility in Safety-Critical Software Systems Development for Aerospace. In: IEEE International Symposium on Software Reliability Engineering Workshops (ISSREW), pp. 222–229 (2022).

\bibitem{Ribeiro2025}
Ribeiro, J., Silva, J., Aguiar, A.: Scrum4DO178C: An Agile Process to Enhance Aerospace Software Development for DO-178C Compliance—A Case Study at Criticality Level A. IEEE Access, vol. 13, pp. 66318-66337 (2025).

\bibitem{Innovation2017}
Kennedy, J.D., Towhidnejad, M.: Innovation and Certification in Aviation Software. In: IEEE Conference on Aviation Software, pp. 1–10 (2017).

\bibitem{ERTS2018}
Marsden, J., Windisch, A., Mayo, R., Grossi, J., Villermin, J., Fabre, L., Aventini, C.:
ED-12C/DO-178C vs. Agile Manifesto—A Solution to Agile Development of Certifiable Avionics Systems. In: ERTS Conference Proceedings (2018).

\bibitem{QScrum2020}
Marques, J.C., Cunha, A.M., Dias, L.A.: Q-Scrum: A Framework for Quality in Safety-Critical Development. In: Communications in Computer and Information Science (CCIS), vol. 1266, pp. 238–245. Springer (2020).

\bibitem{Framework2023}
Baron, C., Louis, V.: Framework and Tooling Proposals for Agile Certification of Safety-Critical Embedded Software in Avionic Systems. Computers in Industry, vol. 148, pp. 103887 (2023). 

\bibitem{A_Lean_and_HighlyAutomated}
Dmitriev, K., Zafar, S.A., Schmiechen, K., Lai, Y., Saleab, M., Nagarajan, P., Dollinger, D., Hochstrasser, M., Holzapfel, F., Myschik, S.:
A Lean and Highly-Automated Model-Based Software Development Process Based on DO-178C/DO-331. In: Proceedings of the IEEE Digital Avionics Systems Conference (DASC) (2020).

\bibitem{Escape_the_waterfall_Ag}
VanderLeest, S.H., Buter, A.: Escape the Waterfall: Agile for Aerospace. In: IEEE Aerospace Conference Proceedings, pp. 1–9 (2009). 

\end{thebibliography}
%

\end{document}